%% file: ECOC_LaTeX_Template.tex
\documentclass[a4paper, oneside, twocolumn, notitlepage, 10pt]{extarticle_ecoc}
\usepackage{ecoc}

\begin{document}
\selectlanguage{english}    % Standard Language

%-------------------------------------------------- Title -----------------------------------------------------%

\title{Record 202.3~Tb/s Transmission over Field-Deployed Fibre using 15.6~THz S+C+L-Bands}%

%------------------------------------------------- Authors-----------------------------------------------------%

\author{
    Jiaqian~Yang\textsuperscript{(1)}, Eric~Sillekens\textsuperscript{(1)}, Benjamin~J.~Puttnam\textsuperscript{(2)}, Ronit~Sohanpal\textsuperscript{(1)}, Mindaugas~Jarmolovičius\textsuperscript{(1)}, \\Romulo~Aparecido\textsuperscript{(1)}, Henrique~Buglia\textsuperscript{(1)}, Ruben~S.~Luis\textsuperscript{(2)}, Ralf~Stolte\textsuperscript{(3)}, Polina~Bayvel\textsuperscript{(1)}, Robert~I.~Killey\textsuperscript{(1)}
}

\maketitle                  % Create title and author

%------------------------------------------ Description of Authors ----------------------------------------------%

\begin{strip}
 \begin{author_descr}
 
   \textsuperscript{(1)} Optical Networks Group, UCL (University College London), London, UK
   \textcolor{blue}{\uline{jiaqian.yang.18@ucl.ac.uk}}\\
   \textsuperscript{(2)} National Institute of Information and Communications Technology, Tokyo, Japan\\
   \textsuperscript{(3)} Coherent / Finisar, New South Wales, Australia

 \end{author_descr}
\end{strip}

\setstretch{1.1}
%-------------------------------------------------- Footnote -------------------------------------------------------%
\renewcommand\footnotemark{}
\renewcommand\footnoterule{}
%\let\thefootnote\relax\footnotetext{text}

%-------------------------------------------------- Abstract ---------------------------------------------------------%

\begin{strip}
  \begin{ecoc_abstract}
    % NOTE: Don't use a blank line here but start abstract right away to avoid an extra line break
    Ultra-wideband, field-deployed metropolitan fibre transmission is experimentally demonstrated, measuring a record 202.3~Tb/s GMI and 189.5~Tb/s after decoding with 20.9~dBm launch power and lumped amplification only. An experimentally-optimised 5~dB pre-tilt over the 15.6~THz optical bandwidth was applied to overcome ISRS. \textcopyright2024 The Author(s)
  \end{ecoc_abstract}
\end{strip}

\setlength{\abovedisplayskip}{3pt}
\setlength{\belowdisplayskip}{3pt}
%-------------------------------------------------- Introduction Section -------------------------------------------------------%

\section{Introduction}
\begin{figure}[htbp]
    \centering
    \input{state_of_art_fig}
    \vspace{-15pt}
    \caption{Throughput and transmission distance of recent deployed fibre experiments. Solid dots: decoded net rate; Hollow dots: rate from GMI.}
    \label{fig:state_of_art}
    \vspace{-5pt}
\end{figure}
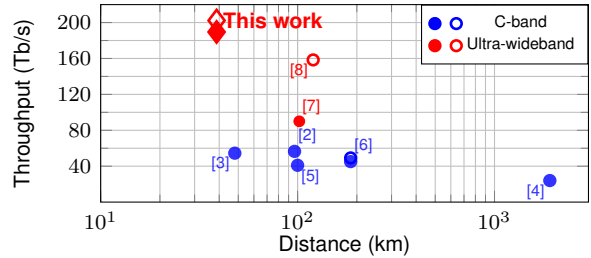

Over the past three decades, optical communication networks have played a vital role in global data transmission~\cite{Winzer:18}. Most field-deployed fibre links use only the C-band with Erbium-doped fibre amplifiers~(EDFA) and similarly the majority of transmission experiments over deployed fibre have been demonstrated over C-band with limited optical bandwidth~\cite{pittala2022_1_71,bajaj2021_54_5,zhang2021field,matsushita2020_41,schuh2020_49}, summarised in Fig.~\ref{fig:state_of_art}. However, S- and L-bands offer a promising route for capacity growth, with low attenuation in standard single-mode fibre (SSMF) and commercially available lumped amplifiers, including Thulium-doped fibre amplifier~(TDFA) for S-band and EDFA for L-band. Several multi-band transmissions over field-deployed fibres have been demonstrated with $>$~100~nm wavelength-division multiplexing (WDM) bandwidth~\cite{renaudier2018field,frignac2023record}, using either semiconductor optical amplifiers with potential high nonlinearity or doped fibre amplifiers with a high launch powers of $>$ 25~dBm. Given the problems that often limit the performance of field-deployed fibres, especially in metropolitan environments, such as fibre ageing and OH contamination, splice and connector losses, heavy traffic-induced mechanical vibration and temperature fluctuation, or safety considerations which frequently limit the total optical power, e.g. to below~20~dBm, it is essential to evaluate the feasibility of high-capacity ultra-wideband transmission outside the ideal laboratory conditions.

In this work, we demonstrated a record transmission throughput (Fig.~\ref{fig:state_of_art}) of 202.3~Tb/s from generalised mutual information (GMI) and 189.5~Tb/s after decoding over 39~km of field-deployed metropolitan fibre in central London, occupying a 125~nm or 15.6~THz optical bandwidth including S-, C- and L-bands, supported by TDFAs and EDFAs. This was achieved through experimentally optimising the launch spectrum, to handle the impact of inter-channel stimulated Raman scattering (ISRS) and balancing the received optical signal-to-noise ratio~(OSNR) and transceiver performance to tackle safety considerations. This result demonstrates the potential of S+C+L band transmission as a practical extension of the capacity in deployed fibre transmission systems with limited power restrictions.
\vspace{-10pt}

\begin{figure*}[htbp]
    \centering
        \includegraphics[width=0.95\linewidth]{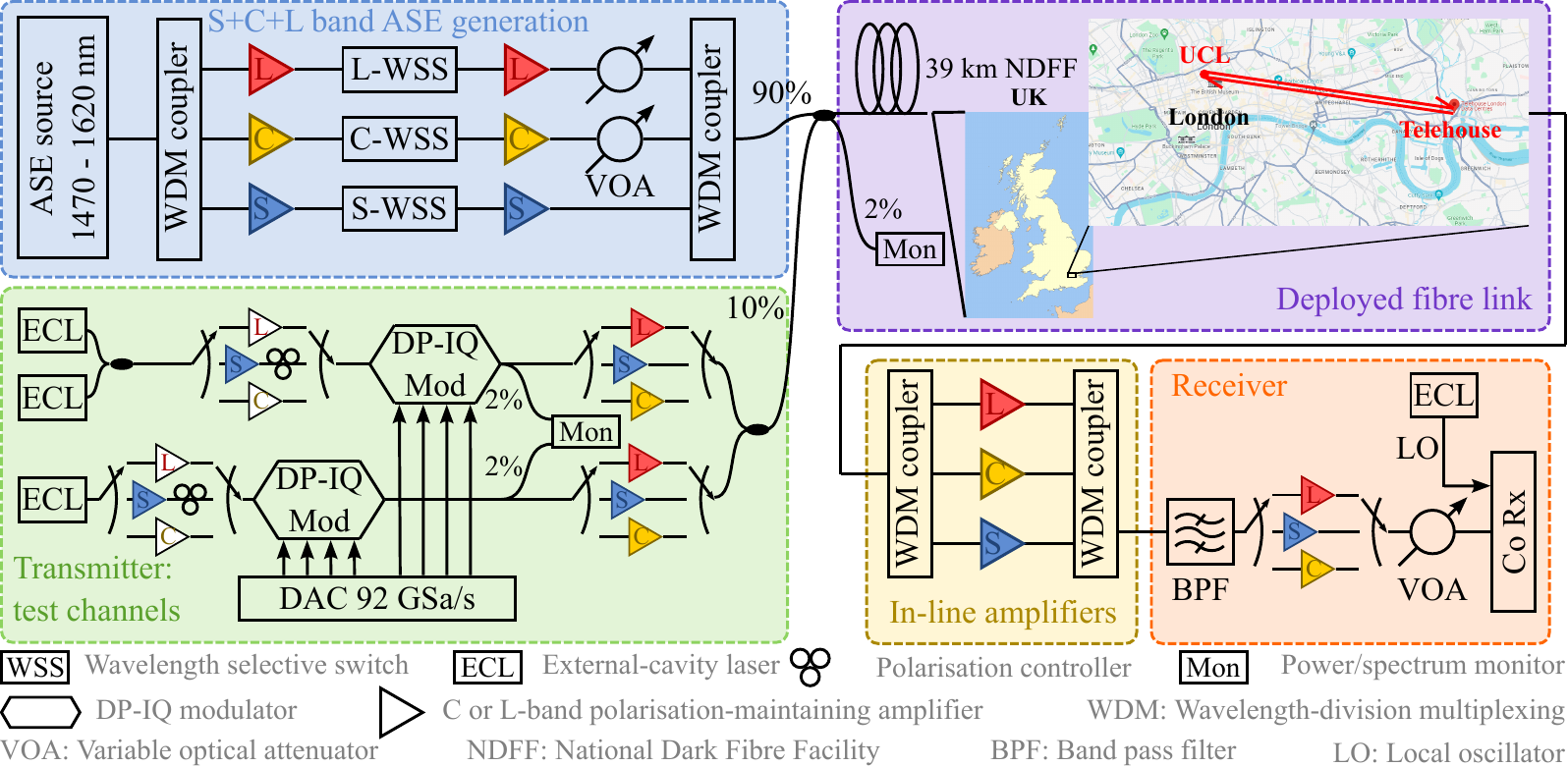}
    \caption{Experimental setup for S-, C-, L-band transmission over 39~km deployed NDFF link.}
    \label{fig:exp}
\end{figure*}

\section{Experimental Demonstration}
Fig.~\ref{fig:exp} shows the experimental setup of the S+C+L band transmission system consisting of 482$\times$ 32~GBaud WDM channels. Three 32.5~GHz-spaced carriers %with \textless100~kHz linewidth 
were generated by tunable lasers and amplified by polarisation maintaining EDFAs in C- and L-bands or TDFAs followed by polarisation alignment in S\nobreakdash-band. The centre channel and the neighbouring channels were modulated with digitally pre-distorted~\cite{geiger2023performance} 32~GBaud root-raised cosine-shaped signals with 1\% roll-off by dual-polarisation in-phase quadrature (DP-IQ) modulators (35~GHz electrical bandwidth), each driven by 92~GSa/s 8-bit digital-to-analogue converters. The three transmitted channels were swept across the S\nobreakdash-, C\nobreakdash-, and L\nobreakdash-bands, with the central channel performance used to assess the system throughput. A variety of geometrically-shaped (GS) constellations were used in each optical band according to the channel signal-to-noise ratio (SNR). Co-propagating channels were emulated using spectrally-shaped amplified spontaneous emission noise (SS-ASE). %~\cite{elson2017investigation}
%between 1480~nm and 1615~nm. 
The SS-ASE was shaped using COHERENT WaveShapers\textsuperscript{\textregistered} which operated as wavelength selective switches (WSSs) to produce a flat or tilted spectrum. A notch was carved by the WSSs to position the three transmitted channels. %Based on the SS-ASE bandwidth and the wavelength of the tunable lasers, the channel-under-test could cover the wavelength range from 1470~nm to 1620 nm. 
%The bandwidth used for the signal was optimised based on the available amplifier output powers and the achievable SNR.

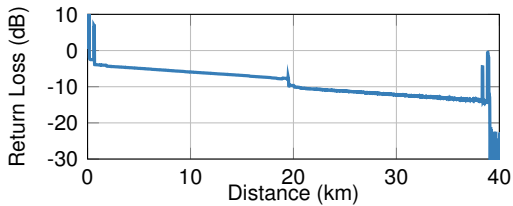
\begin{figure}
    \centering
    \begin{tikzpicture}[font=\footnotesize]
        \begin{axis}[
            width=7cm,
            height=3.5cm,
            %title style={align=left},
            %title={Throughput over\\deployed fibre link},
            %xmode=log,
            xlabel={Distance (km)},
            ylabel={Return Loss (dB)},
            x label style={at={(axis description cs:0.5,-0.15)},anchor=north, yshift = 17},
            y label style={at={(axis description cs:-0.1,.5)},anchor=south, yshift=-32pt},
            grid=both,
            %minor y tick num=1,
            ymax=10,
            ymin=-30,
            xmin = 0,
            xmax = 40,
            xtick={0,10,20,30,40},
            xticklabels={0,10,20,30,40},
            ytick={-30,-20,-10,0,10},
            yticklabels={-30,-20,-10,0,10},
            %nodes near coords,
            %nodes near coords style={
                %font=\footnotesize,
                %inner sep=1pt,
             %},
            %reverse legend,
            %clip=false,
        ]
        \addplot[Set1-B,line width=1.2pt] table[x=distance,y=rl] {OTDR.txt};
        \end{axis}
    \end{tikzpicture}
    \caption{Return loss along the link measured by OTDR.}
    \label{fig:rl}
    \vspace{-10pt}
\end{figure}

The transmission was carried out over a field-deployed fibre link, part of EPSRC National Dark Fibre Facility (NDFF), between UCL (University College London) and the Telehouse datacentre in East London, UK, where a switch looped the connection back to UCL. The total distance is 39~km with a loss of 15~dB (7.8~dB fibre loss and an additional 7.2~dB splicing loss and insertion loss). The attenuation along the link was measured using an optical time domain reflectometer (OTDR) and is shown in Fig.~\ref{fig:rl}. %A 2~\%~tap for power spectrum monitoring was used before the link to balance the power density of test channels and co-propagating ASE. 
The total launch power at UCL was restricted to 20.9~dBm, but, due to high insertion losses, the launch power into the NDFF fibre span was only 17~dBm, with 13.9~dBm in S-band, 11.3~dBm in C-band, and 10.8~dBm in L-band.

In-line TDFA and EDFAs were positioned after fibre transmission, and the %received channel under test was selected using an optical bandpass filter% with a tunable centre wavelength and a fixed bandwidth of 35~GHz
%The residual interfering channels were filtered out in the digital signal processing (DSP). 
%The channel power was boosted by a pre-amplifier in the corresponding band, and a 70-GHz coherent receiver detected the DP-IQ signal. A 10~bit 256~GSa/s real-time oscilloscope was used to digitise and capture the waveform, and details of the pilot-based DSP can be found in \cite{wakayama20212048}. 
receiver consisted of an optical bandpass filter, a pre-amplifier in the corresponding band, a 70-GHz coherent receiver, and a 10~bit 256~GSa/s real-time oscilloscope. Off-line pilot-based DSP~\cite{wakayama20212048} was operated on 2 samples per symbol. System performance was quantified in terms of SNR, and the achievable information rate (AIR) after deducting the 4.64~\% pilot overhead was estimated from GMI for all 482 channels. DVB-S2X low-density parity check codes were used to decode the received symbols \cite{dvb} with adaptive code rate, and code-rate puncturing was applied to achieve a bit error rate below 3$\times$10\textsuperscript{-4} with a 0.5~\% outer hard-decision code overhead.

\begin{figure}[b!]
\centering
    \begin{subfigure}{\linewidth}
        \begin{tikzpicture}[font=\footnotesize]
        \begin{axis}[
        width=7.8cm,
        height=3cm,
        ylabel={SNR (dB)},
        xmin=1480, xmax=1530,
        ymin=16, ymax=20,
        /pgf/number format/1000 sep={},
        xtick={1480,1505,1530},
        xticklabels={1480,1505,1530},
        ytick={17,18,19},
        yticklabels={17,18,19},
        grid=both,
        %xlabel near ticks,
        %ylabel near ticks,
        x label style={at={(axis description cs:.5,.10)}},
        y label style={at={(axis description cs:0.08,.5)}},
        %legend pos=north east,
        ylabel style={font=\fontsize{6}{0}\selectfont},
        legend style={font=\fontsize{6}{0}\selectfont, row sep=-1.5pt, at={(1.2,1)}},
        ]
        \node[anchor=west] at (rel axis cs:.9,.85){\footnotesize(a)};
        \addplot[Set1-B,line width=1.2pt] table[x=wavelength,y=OSNR] {Notch_S_tilt5_Rx.txt};
        \addplot[Set1-A,mark=*,only marks, mark size=1.5, mark options={solid}] table[x=wav_notches,y=OSNR] {Notch_S_tilt5_Rx_notches.txt};
        \addplot[Set1-E,dashed,line width=1.2pt] table{
        1480    19
        1526    19
        };
        \end{axis}
        \end{tikzpicture}
        % \vspace{-.6em}
        % \caption{}
        % \label{fig:OSNR_S}
    \end{subfigure}
    \begin{subfigure}{\linewidth}
        \begin{tikzpicture}[font=\footnotesize]
        \begin{axis}[
        width=7.8cm,
        height=3cm,
        ylabel={SNR (dB)},
        xmin=1530, xmax=1570,
        ymin=18, ymax=26,
        /pgf/number format/1000 sep={},
        xtick={1530,1550,1570},
        xticklabels={1530,1550,1570},
        ytick={20,22,24},
        yticklabels={20,22,24},
        grid=both,
        %xlabel near ticks,
        %ylabel near ticks,
        x label style={at={(axis description cs:.5,.10)}},
        y label style={at={(axis description cs:0.08,.5)}},
        %legend pos=north east,
        ylabel style={font=\fontsize{6}{0}\selectfont},
        legend style={font=\fontsize{6}{0}\selectfont, row sep=-1.5pt, at={(1.2,1)}},
        ]
        \node[anchor=west] at (rel axis cs:.9,.85){\footnotesize(b)};
        \addplot[Set1-B,line width=1.2pt] table[x=wavelength,y=OSNR] {Notch_C_tilt5_Rx.txt};
        \addplot[Set1-A,mark=*,only marks, mark size=1.5, mark options={solid}] table[x=wav_notches,y=OSNR] {Notch_C_tilt5_Rx_notches.txt};
        \addplot[Set1-E,dashed,line width=1.2pt] table{
        1530    23
        1566    23
        };
        \end{axis}
        \end{tikzpicture}
    % \caption{}
    % \label{fig:OSNR_C}
        % \vspace{-.6em}
        \end{subfigure}
    \begin{subfigure}{\linewidth}
        \begin{tikzpicture}[font=\footnotesize]
        \begin{axis}[
        width=7.8cm,
        height=3cm,
        xlabel={Wavelength (nm)},
        ylabel={SNR (dB)},
        xmin=1570, xmax=1620,
        ymin=20, ymax=25,
        /pgf/number format/1000 sep={},
        xtick={1570,1595,1620},
        xticklabels={1570,1595,1620},
        ytick={21,22,23},
        yticklabels={21,22,23},
        grid=both,
        %xlabel near ticks,
        %ylabel near ticks,
        x label style={at={(axis description cs:.5,.10)}},
        y label style={at={(axis description cs:0.08,.5)}},
        %legend pos=north east,
        ylabel style={font=\fontsize{6}{0}\selectfont},
        legend style={font=\fontsize{6}{0}\selectfont, row sep=-1.5pt, at={(1,1)}, legend columns=3},
        ]
        \node[anchor=east] at (rel axis cs:1,.5){\footnotesize(c)};
        \addplot[Set1-A,mark=*,only marks, mark size=1.5, mark options={solid}] table[x=wav_notches,y=OSNR] {Notch_L_tilt5_Rx_notches.txt};\addlegendentry{Notch est. SNR}
        \addplot[Set1-B,line width=1.2pt] table[x=wavelength,y=OSNR] {Notch_L_tilt5_Rx.txt};\addlegendentry{SNR interp.}
        \addplot[Set1-E,dashed,line width=1.2pt] table{
        1570    21
        1620    21
        };\addlegendentry{Transceiver lim.}
        
        \end{axis}
        \end{tikzpicture}
    % \caption{}
    % \label{fig:OSNR_L}
    \end{subfigure}
    \caption{(a)-(c) SNR estimated from notch sweeping for S-, C-, and L-band respectively (dots: measurement points, solid line: interpolation to all WDM channels) and transceiver-limited SNR (dashed line).}
    \label{fig:notch}
\end{figure}
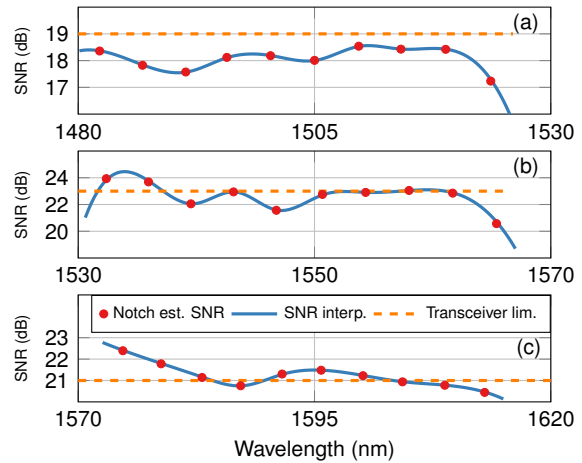

%The launch power spectrum was optimised prior to transmission. 
Since the signal quality is affected by the joint effect of transceiver SNR and the ASE noise floor in the notch carved by the WSS, we first measured the back-to-back SNR for each channel and estimated the received SNR after fibre transmission and lumped amplification with the Gaussian noise model~\cite{semrau2019closed}. This yielded the received SNR values 19~dB, 23~dB, and 21~dB, for S\nobreakdash-, C\nobreakdash-, and L\nobreakdash-band channels, respectively. The launch power tilt was optimised by carving 10 notches in each band and measuring the achievable OSNR using the received noise floor within the notches. This OSNR could be interpolated to all channels and mapped to SNR and capacity using Shannon's equation. The pre-tilt of the launch power spectrum was gradually increased to reduce the ISRS power transfer from S- to L-band in the optimisation process, allowing the increase of usable optical bandwidth of S- and L-bands.

Fig.~\ref{fig:notch}(a)-(c) shows the interpolated SNR estimated from the notch sweeping and the transmission-limited SNR across the three bands. The S-band received SNR was the lowest among the three bands. This was due to the limited output power of the S-band TDFA. Taking the estimated throughput as a metric, the best overall performance was seen with 5~dB pre-tilt and bandwidths of 1480-1526~nm, 1530-1566~nm, 1572-1615~nm in the 3 bands, resulting in 187$\times$, 140$\times$, and 155$\times$ channels, respectively. With this configuration, the estimated total throughput was 208.8~Tb/s.

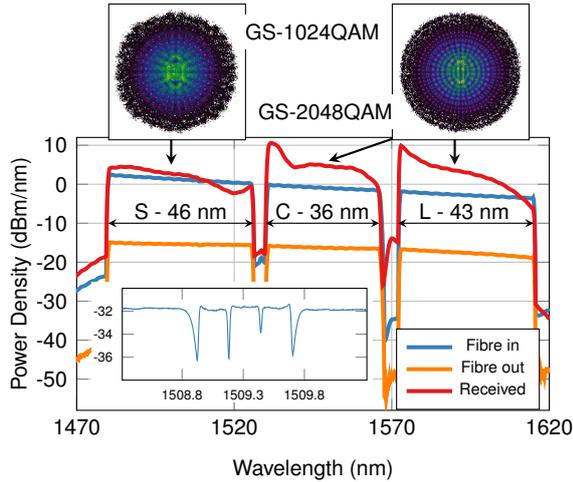
\begin{figure}
    \centering
        \begin{tikzpicture}[font=\footnotesize]

    \begin{axis}[
    width=7.8cm,
    height=5.2cm,
    xlabel={Wavelength (nm)},
    ylabel={Power Density (dBm/nm)},
    xmin=1470, xmax=1620,
    ymin=-58, ymax=12,
    /pgf/number format/1000 sep={},
    xtick={1470,1520,1570,1620},
    xticklabels={1470,1520,1570,1620},
    ytick={-50,-40,-30,-20,-10,0,10},
    yticklabels={-50,-40,-30,-20,-10,0,10},
    grid=both,
    %xlabel near ticks,
    %ylabel near ticks,
    x label style={at={(axis description cs:.5,.00)}},
    y label style={at={(axis description cs:0.08,.5)}},
    %legend pos=north east,
    %ylabel style={font=\fontsize{6}{0}\selectfont},
    legend style={font=\fontsize{6}{0}\selectfont, row sep=-1.5pt, at={(0.98,0.30)}}
    ]

    %\node[anchor=west] at (rel axis cs:.9,.93){\footnotesize(d)};
    \addplot[Set1-B,line width=1.5pt] table[x=wavelength,y=fibrein] {Spectrum_RBW0.745nm.txt};\addlegendentry{Fibre in}
    \addplot[Set1-E,line width=1.5pt] table[x=wavelength,y=fibreout] {Spectrum_RBW0.745nm.txt};\addlegendentry{Fibre out}
    \addplot[Set1-A,line width=1.5pt] table[x=wavelength,y=rx] {Spectrum_RBW0.745nm.txt};\addlegendentry{Received}
    \draw[>=stealth, <->](axis cs:1480,-10)--(axis cs:1526,-10);
    \draw[>=stealth, <->](axis cs:1530,-10)--(axis cs:1566,-10);
    \draw[>=stealth, <->](axis cs:1572,-10)--(axis cs:1615,-10);
    \node at (axis cs: 1503,-7) {S - 46~nm};
    \node at (axis cs: 1548,-7) {C - 36~nm};
    \node at (axis cs: 1593,-7) {L - 43~nm};

    \draw[>=stealth, ->, thick] (axis cs:1500,20)--(axis cs:1500,5);
    \draw[>=stealth, ->, thick] (axis cs:1600,20)--(axis cs:1550,6);
    \draw[>=stealth, ->, thick] (axis cs:1590,20)--(axis cs:1590,5);
    \end{axis}

    \node[inner sep=0pt, anchor=west, draw=black, fill=white] (qam1) at (rel axis cs:0,1.24) {\includegraphics[width=1.8cm]{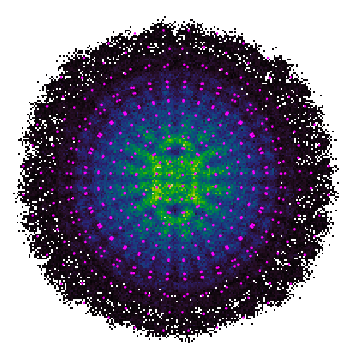}};
    \node[inner sep=0pt, anchor=west, draw=black, fill=white] (qam2) at (rel axis cs:.6,1.24) {\includegraphics[width=1.8cm]{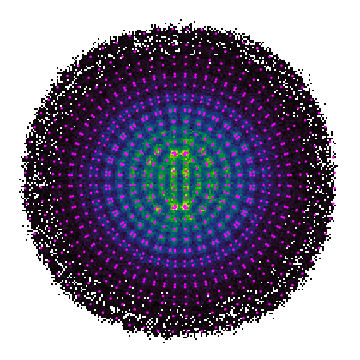}};
    \node[anchor=west,xshift=2.2em,yshift=1.5em,fill=white] at (qam1){GS-1024QAM};
    \node[anchor=east,xshift=-2.2em,yshift=-1.5em,fill=white] at (qam2){GS-2048QAM};
    
    \fill [white] (0.2,0.1) rectangle ++(3.7,1.6);
    \begin{axis}[
    xshift=0.6cm,
    yshift=0.4cm,
    width=4.8cm,
    height=2.8cm,
    xmin=1508.311, xmax=1510.311,
    ymin=-38, ymax=-30,
    xtick={1508.8,1509.3,1509.8},
    xticklabels={1508.8,1509.3,1509.8},
    ytick={-36,-34,-32},
    yticklabels={-36,-34,-32},
    xticklabel style={/pgf/number format/1000 sep=},
    %grid=both,
    %xlabel near ticks,
    %ylabel near ticks,
    %y label style={at={(axis description cs:.3,.5)}},
    font=\tiny,
    opacity=1,
    ]
    \addplot[Set1-B] table[x=wavelength,y=fibrein]{Spectrum_3CH_RBW0.018nm.txt};
    \end{axis}
    \end{tikzpicture}
    \caption{Signal power spectra at the fibre input, fibre output and receiver. Inset: modulated 3 signal channels positioned within the notch.}
    \label{fig:result}
\end{figure}

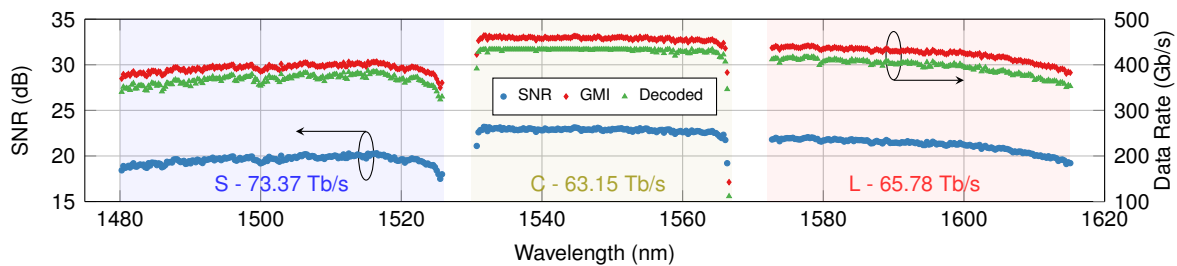
\begin{figure*}[b!]
    \centering
        \begin{tikzpicture}[font=\footnotesize]
            \begin{axis}[
                width=15cm,
                height=4cm,
                axis y line*=left,
                xlabel={Wavelength (nm)},
                ylabel={SNR (dB)},
                xmin=1475, xmax=1620,
                ymin=15, ymax=35,
                xtick={1480,1500,1520,1540,1560,1580,1600,1620},
                xticklabels={1480,1500,1520,1540,1560,1580,1600,1620},
                ytick={15,20,25,30,35},
                yticklabels={15,20,25,30,35},
                grid=both,
                xticklabel style={/pgf/number format/1000 sep=},
                xlabel near ticks,
                %ylabel near ticks,
                y label style={at={(axis description cs:.03,.5)}},
                legend pos=north east,
                legend columns=2,
                legend style={font=\fontsize{6}{0}\selectfont, row sep=-2.5pt},
                legend cell align={left},
            ]
            \addplot[Set1-B,only marks,mark=*,mark options={solid,Set1-B},mark size=1] table[x=wavelength,y=SNR]{SNR_GMI.txt};
            \node(sources) at (axis cs:1480,15){};
            \node(destinations) at (axis cs:1526,35){};
            \node(sourcec) at (axis cs:1530,15){};
            \node(destinationc) at (axis cs:1567,35){};
            \node(sourcel) at (axis cs:1572,15){};
            \node(destinationl) at (axis cs:1615,35){};
            \draw[fill=blue,opacity=0.05,draw=none] (sources) rectangle (destinations);
            \draw[fill=olive,opacity=0.05,draw=none] (sourcec) rectangle (destinationc);
            \draw[fill=red,opacity=0.05,draw=none] (sourcel) rectangle (destinationl);
            \node[blue,opacity=0.8] at (axis cs: 1503,17) {S - 73.37~Tb/s};
            \node[olive,opacity=0.8] at (axis cs: 1548,17) {C - 63.15~Tb/s};
            \node[red,opacity=0.8] at (axis cs: 1593,17) {L - 65.78~Tb/s};
            \draw (axis cs:1515,20) ellipse (0.1cm and 0.32cm);
            \draw (axis cs:1590,31) ellipse (0.1cm and 0.32cm);
            \draw[>=stealth, ->](axis cs:1515,22.7)--(axis cs:1505,22.7);
            \draw[>=stealth, ->](axis cs:1590,28.3)--(axis cs:1600,28.3);
            \end{axis}
            \begin{axis}[
                width=15cm,
                height=4cm,
                axis y line*=right,
                axis x line=none,
                ylabel={Data Rate (Gb/s)},
                xmin=1475, xmax=1620,
                ymin=100, ymax=500,
                ytick={100,200,300,400,500},
                yticklabels={100,200,300,400,500},
                %xlabel near ticks,
                %ylabel near ticks,
                y label style={at={(axis description cs:1.15,.5)}},
                legend pos=north east,
                legend columns=3,
                legend style={font=\fontsize{6}{0}\selectfont, column sep=2.5pt, at={(0.62,0.68)}},
                legend cell align={left},
            ]
            \addlegendentry{SNR};
            \addlegendimage{Set1-B,only marks,mark=*,mark options={solid,Set1-B},mark size=1}
            \addlegendentry{GMI};
            \addplot[Set1-A,only marks,mark=diamond*,mark size=1] table[x=wavelength,y=DataRate]{SNR_GMI.txt};
            \addlegendentry{Decoded};
            \addplot[Set1-C,only marks,mark=triangle*,mark size=1] table[x=wavelength,y=NetRate]{SNR_GMI.txt};
            \end{axis}
        \end{tikzpicture}
    %\vspace{-3pt}
    \caption{Received SNR and data rate per channel estimated from GMI and after decoding in the deployed NDFF transmission with 187$\times$S-, 140$\times$C- and 155$\times$L-band channels.}
    \label{fig:SNRGMI}
\end{figure*}
\vspace{-5pt}

\section{Results and Discussion}
Fig.~\ref{fig:result} shows the measured tilted spectrum at the fibre input with the inset showing the 3$\times$32~GHz channels placed within the sliding notch. The effect of wavelength-dependent fibre loss and ISRS after 39~km deployed fibre transmission is shown by the orange line, with approximately 17~dB uniform loss in the 3 bands. The red line shows the received spectrum after amplification, where the received signal power is recovered to values similar to the launch powers. %The trade-off between the S- and L-band power spectral density requires launch power tilt optimisation. Fig.~\ref{fig:result}a-\ref{fig:result}c shows the received OSNR measured from the notch for different wavelengths per band. 
%The dashed lines show the transceiver SNR, balancing the OSNR and transceiver performance allowed us to increase throughput by re-allocation of the transmission power to worse-performing channels. 
The figure also shows the geometrically-shaped modulation formats transmitted for each band, optimised for the channel SNR, GS-1024~QAM and GS-2048~QAM for 19-dB and 22-dB SNR respectively.  \cite{9852813}. The GS-2048~QAM was used in the C-, and L-bands, and the GS-1024~QAM was used in the S-band.%Such allocation was given considering the achieved SNR with a rectangular QAM configuration, and the choice for the shorter half of S-band and L-band is GS-64 QAM 12 dB and GS-256 QAM 16dB for the longer S-band half and C-band.

Fig.~\ref{fig:SNRGMI} shows the received signal measurements, with the SNR, GMI, and decoded data rate, after the transmission over 39~km. The total optical bandwidth used was 15.6~THz.
Throughputs of 73.37~Tb/s, 63.15~Tb/s, and 65.78~Tb/s were achieved for S-, C-, and L-band, respectively, totalling 202.3~Tb/s from GMI and 189.5~Tb/s after decoding. Despite supporting a lower data rate per channel in S-band, a greater number of carriers between 1480~and 1520~nm makes this band the largest contributor to throughput. The L-band can extend the bandwidth to longer wavelengths, contributing 155 channels to the total. The C-band fits only 140 channels, but these channels have the highest data rate per channel. The combination of these 3 bands allowed us to transmit a record 202.3~Tb/s over field-deployed fibre.

\vspace{-5pt}

\section{Conclusions}
A record throughput over field-deployed fibre using an ultra-wideband (15.6~THz) S+C+L-band transmission system was experimentally demonstrated with the total data rate of 202.3~Tb/s from GMI and 189.5~Tb/s after decoding. The launch power profile was optimised to overcome the inter-channel stimulated Raman scattering, balancing OSNR and transceiver performance. The combination of system optimisation, constellation shaping and state-of-the-art DSP resulted in this record-breaking demonstration, showing that S+C+L-band is a practical solution to increase the capacity of deployed fibre transmission systems.

\section{Acknowledgements}
This work was supported by EPSRC grants EP/R035342/1 TRANSNET (Transforming networks - building an intelligent optical infrastructure), EP/W015714/1 EWOC (Extremely Wideband Optical Fibre Communication Systems), EP/S028854/1 NDFF (National Dark Fibre Facility), EP/V007734/1 EPSRC Strategic equipment grant, EP/T517793/1 EPSRC studentship, and Microsoft Research.% Does NDFF need to be acknowledged?

\printbibliography

\vspace{-4mm}

%%%%%%%%%%%%%%%%%%%%%%%%%%%%%%%%%%%%%%%%%%%%%
%---------------------------------------------- End of Document -----------------------------------------------%
\end{document}

%% file: state_of_art_fig.tex
\begin{tikzpicture}[font=\footnotesize]
\begin{axis}[
    width=8cm,
    height=4.2cm,
    xmode=log,
    xlabel={Distance (km)},
    ylabel={Throughput (Tb/s)},
    x label style={at={(axis description cs:0.5,-0.15)},anchor=north, yshift = 17},
    y label style={at={(axis description cs:-0.1,.5)},anchor=south, yshift=-32pt},
    grid=both,
    minor y tick num=1,
    ymax=220,
    ymin=0,
    xmin = 10,
    xmax = 3000,
    xtick={10,100,1000},
    ytick={40,80,120,160,200},
    %nodes near coords,
    %nodes near coords style={
        %font=\footnotesize,
        %inner sep=1pt,
    %},
    reverse legend,
    clip=false,
    legend style={font=\fontsize{6}{0}\selectfont, row sep=-1.5pt, at={(1,1)}, legend columns=2},
]

%\addplot[Set1-D, dashed, line width=1.2pt] table[x=dis,y=thr] {Unrepeated_223km_LR.txt};
%\node[Set1-D] at (axis cs: 380,80) {$25\text{Pb/s}\cdot\text{km}$};

\newcommand{\datapointfec}[3]{
    \addplot[only marks, mark=*, mark size=2.0pt, color=red,
        point meta=explicit symbolic, 
        nodes near coords={\cite{#1}},
        nodes near coords style={
            font=\footnotesize,
            inner sep=1pt,
            anchor=#3,
        }, forget plot,
    ] coordinates {#2};
}

\newcommand{\datapointgmi}[3]{
    \addplot[only marks, mark=o, mark size=2.0pt, color=red, line width=1pt,
        point meta=explicit symbolic, 
        nodes near coords={\cite{#1}},
        nodes near coords style={
            font=\footnotesize,
            inner sep=1pt,
            anchor=#3,
        }, forget plot,
    ] coordinates {#2};
}

\newcommand{\datapointfecc}[3]{
    \addplot[only marks, mark=*, mark size=2.0pt, color=blue, line width=1pt, opacity=0.8,
        point meta=explicit symbolic, 
        nodes near coords={\cite{#1}},
        nodes near coords style={
            font=\footnotesize,
            inner sep=1pt,
            anchor=#3,
        }, forget plot,
    ] coordinates {#2};
}

\newcommand{\datapointgmic}[3]{
    \addplot[only marks, mark=o, mark size=2.0pt, color=blue, line width=1pt,
        point meta=explicit symbolic, 
        nodes near coords style={
            font=\footnotesize,
            inner sep=1pt,
            anchor=#3,
        }, forget plot,
    ] coordinates {#2};
}

%\node[anchor=west] at (rel axis cs:.8,.95){\footnotesize(a)};
\datapointfecc{pittala2022_1_71}{(96.5,56.51)}{south west}
\datapointfecc{bajaj2021_54_5}{(48,54.5)}{north east}
\datapointfecc{zhang2021field}{(1910,24)}{north east}
\datapointfecc{matsushita2020_41}{(100,41)}{north west}
\datapointgmic{schuh2020_49}{(186,49.2)}{south west}
\datapointfecc{schuh2020_49}{(186,45.1)}{south west}
\datapointfec{renaudier2018field}{(102,90)}{south west}
\datapointgmi{frignac2023record}{(120,158.4)}{north east}

% Custom marker for "Our work" point (red triangle)
\addplot[line width=1pt,only marks,mark=diamond, mark options={mark size = 4, red},
    point meta=explicit symbolic, 
    nodes near coords={\textbf{This work}},
    nodes near coords style={
        font=\footnotesize,
        inner sep=2pt,
        anchor=west,
    },
] coordinates {(38.8,202.3)};

\addplot[line width=1pt,only marks,mark=diamond*, mark options={mark size = 4, red, fill=red},
    point meta=explicit symbolic, 
] coordinates {(38.8,189.5)};

\addlegendentry{Ultra-wideband}
\addlegendimage{only marks, mark=o, mark size=2.0pt, color=red, line width=1pt}
\addlegendentry{}
\addlegendimage{only marks, mark=*, mark size=2.0pt, color=red, line width=1pt}
\addlegendentry{C-band}
\addlegendimage{only marks, mark=o, mark size=2.0pt, color=blue, line width=1pt}
\addlegendentry{}
\addlegendimage{only marks, mark=*, mark size=2.0pt, color=blue, line width=1pt}

%\node[fill=white, inner sep=0pt, above left] at (axis cs: 10,220) {(Tb/s)};

\end{axis}
\end{tikzpicture}

%% file: references.bib
@article{Winzer:18,
author = {Peter J. Winzer and David T. Neilson and Andrew R. Chraplyvy},
journal = {Optics Express},
number = {18},
pages = {24190--24239},
publisher = {Optica Publishing Group},
title = {Fiber-optic transmission and networking: the previous 20 and the next 20 years [Invited]},
volume = {26},
month = {Sep},
year = {2018},
doi = {10.1364/OE.26.024190},
}

@ARTICLE{pittala2022_1_71,
  author={Pittalà, Fabio and Braun, Ralf-Peter and Böcherer, Georg and Schulte, Patrick and Schaedler, Maximilian and Bettelli, Stefano and Calabrò, Stefano and Kuschnerov, Maxim and Gladisch, Andreas and Westphal, Fritz-Joachim and Xie, Changsong and Chen, Rongfu and Wang, Qibing and Zheng, Bofang},
  journal={IEEE Photonics Technology Letters}, 
  title={1.71 {Tb/s} Single-Channel and 56.51 {Tb/s DWDM} Transmission Over 96.5 km Field-Deployed {SSMF}}, 
  year={2022},
  volume={34},
  number={3},
  pages={157-160},
  doi={10.1109/LPT.2022.3142538}}

@inproceedings{bajaj2021_54_5,
author = {Vinod Bajaj and Fred Buchali and Mathieu Chagnon and Sander Wahls and Vahid Aref},
booktitle = {Optical Fiber Communication Conference},
journal = {Optical Fiber Communication Conference},
pages = {M5F.2},
publisher = {Optica Publishing Group},
title = {54.5 {Tb/s WDM} Transmission over Field Deployed Fiber Enabled by Neural Network-Based Digital Pre-Distortion},
year = {2021},
doi = {10.1364/OFC.2021.M5F.2},
}

@article{zhang2021field,
author = {Anxu Zhang and Junjie Li and Lipeng Feng and Kai Lv and Fei Yan and Yusen Yang and Haiqiang Wang and Qifang Yang and Lingquan Wang and Xiaolei Zhang and Shibao Ding and Min Liao and Yi Yu and Liangchuan Li},
journal = {Optics Express},
number = {26},
pages = {43811--43818},
publisher = {Optica Publishing Group},
title = {Field trial of 24-{Tb/s} (60 × 400{Gb/s) DWDM} transmission over a 1910-km {G.654.E} fiber link with 6-{THz}-bandwidth {C-band EDFAs}},
volume = {29},
year = {2021},
doi = {10.1364/OE.447553},
}

@article{matsushita2020_41,
author = {Asuka Matsushita and Masanori Nakamura and Shuto Yamamoto and Fukutaro Hamaoka and Yoshiaki Kisaka},
journal = {Journal of Lightwave Technology},
number = {11},
pages = {2905--2911},
publisher = {Optica Publishing Group},
title = {{41-Tbps C}-Band {WDM} Transmission With {10-bps/Hz} Spectral Efficiency Using 1-{Tbps/}$\lambda$ Signals},
volume = {38},
year = {2020},
}

@INPROCEEDINGS{schuh2020_49,
  author={Schuh, Karsten and Buchali, Fred and Dischler, Roman and Chagnon, Mathieu and Aref, Vahid and Buelow, Henning and Hu, Qian and Pulka, Florian and Frascolla, Massimo and Alhammadi, Esmaeel and Samhan, Adel and Younis, Islam and El-Zonkoli, Mohamed and Winzer, Peter},
  booktitle={Optical Fiber Communications Conference and Exhibition}, 
  title={49.2-{Tbit/s WDM} Transmission over 2×93-km Field-Deployed Fiber}, 
  year={2020},
  pages={M4K.2},
  doi={}}

@INPROCEEDINGS{renaudier2018field,
  author={Renaudier, J. and Meseguer, A. Carbo and Ghazisaeidi, A. and Brindel, P. and Tran, P. and Verdier, A. and Blache, F. and Makhsiyan, M. and Mekhazni, K. and Calo, C. and Debregeas, H. and Pulka, F. and Suignard, J. and Boutin, A. and Fontaine, N. and Neilson, D. and Ryf, R. and Chen, H. and Dellinger, R. and Grubb, S. and Achouche, M. and Charlet, G.},
  booktitle={European Conference on Optical Communication}, 
  title={Field Trial of 100nm Ultra-Wideband Optical Transport with {42GBd 16QAM} Real-Time and {64GBd PCS64QAM} Channels}, 
  year={2018},
  pages={Th1D.6},
  doi={10.1109/ECOC.2018.8535349}}

@INPROCEEDINGS{frignac2023record,
  title={{Record 158.4 Tb/s transmission over 2x60 km field SMF using {S+ C+ L 18THz}-bandwidth lumped amplification}},
  author={Frignac, Yann and Le Gac, Dylan and Lorences-Riesgo, Abel and Godard, Loig and Landero, Salma Escobar and Zhao, Xiaohui and Pincemin, Erwan and Le Guyader, Bertrand and Brochier, Nicolas and Guo, Qiang and others},
  year={2023},
  booktitle={European Conference on Optical Communications}, 
  year={2023},
  pages={M.A.5.2},
  doi={10.1049/icp.2023.2242}
}

@ARTICLE{geiger2023performance,
  author={Geiger, Benedikt and Sillekens, Eric and Ferreira, Filipe and Killey, Robert and Galdino, Lidia and Bayvel, Polina},
  journal={Journal of Lightwave Technology}, 
  title={On the Performance Limits of High-Speed Transmission Using a Single Wideband Coherent Receiver}, 
  year={2023},
  volume={41},
  number={12},
  pages={3816-3824},
  doi={10.1109/JLT.2023.3277624}}

@article{wakayama20212048,
author = {Yuta Wakayama and Thomas Gerard and Eric Sillekens and L\'{i}dia Galdino and Domani\c{c} Lavery and Robert I. Killey and Polina Bayvel},
journal = {Optics Express},
number = {12},
pages = {18743--18759},
publisher = {Optica Publishing Group},
title = {2048-{QAM} transmission at 15 {GBd} over 100 km using geometric constellation shaping},
volume = {29},
month = {Jun},
year = {2021},
doi = {10.1364/OE.423361},
}

@ARTICLE{semrau2019closed,
  author={Semrau, Daniel and Killey, Robert I. and Bayvel, Polina},
  journal={Journal of Lightwave Technology}, 
  title={A Closed-Form Approximation of the {Gaussian} Noise Model in the Presence of Inter-Channel Stimulated {Raman} Scattering}, 
  year={2019},
  volume={37},
  number={9},
  pages={1924-1936},
  doi={10.1109/JLT.2019.2895237}}

@ARTICLE{9852813,
  author={Sillekens, Eric and Liga, Gabriele and Lavery, Domaniç and Bayvel, Polina and Killey, Robert I.},
  journal={Journal of Lightwave Technology}, 
  title={High-Cardinality Geometrical Constellation Shaping for the Nonlinear Fibre Channel}, 
  year={2022},
  volume={40},
  number={19},
  pages={6374-6387},
  doi={10.1109/JLT.2022.3197366}}

@inproceedings{dvb,
  author={{Digital Video Broadcasting (DVB)}},
  title={Second Generation Framing Structure, Channel Coding and Modulation Systems for Broadcasting, Interactive Services, News Gathering and Other Broadband Satellite Applications; {Part 2: DVB-S2} Extensions {(DVB-S2X)}}, 
  year={2021},
  booktitle={ETSI Standard SIST EN 302 307-2 V1.2.1:2020, European Telecommunications Standards Institute (ETSI), Sophia-Antipolis, France}
}
